\documentclass[preprint, superscriptaddress,amsmath, nofootinbib]{revtex4-1}
\usepackage{graphicx}
\usepackage{latexsym}
\usepackage{slashed}
\usepackage{bm}
\usepackage{color}
\usepackage[colorlinks,citecolor=blue,urlcolor=blue,linkcolor=red]{hyperref}
\usepackage{diagbox}
\usepackage{makecell}
\usepackage{hyperref}
\pdfoutput=1
\usepackage{dcolumn}% Align table columns on decimal point
\usepackage{bm}% bold math
\usepackage{multirow}
\usepackage{subcaption}
\usepackage{ulem}
\usepackage{tabularx}
\usepackage{hyperref}
\usepackage{cleveref}
\usepackage{xcolor} % 

\newcommand{\Eq}[1]{Eq.~(\ref{#1})}

\definecolor{CJ}{RGB}{128, 0, 128}

\begin{document}

\title{Probing Double-Peaked Gamma-Ray Spectra from Primordial Black Holes with Next-Generation Gamma-Ray Experiments}
\def\slash#1{#1\!\!\!/}

\author{C.J. Ouseph}
\email{ouseph444@gmail.com}
\affiliation{Institute of Convergence Fundamental Studies, Seoul National University
of Science and Technology, Seoul 01811, Korea}

\author{Giorgio Busoni}
\email{giorgio.busoni@adelaide.edu.au}
\affiliation{ARC Centre of Excellence for Dark Matter Particle Physics, Department of Physics, University
of Adelaide, South Australia 5005, Australia} 

\author{John Gargalionis}
\email{john.gargalionis@adelaide.edu.au}
\affiliation{ARC Centre of Excellence for Dark Matter Particle Physics, Department of Physics, University
of Adelaide, South Australia 5005, Australia} 

\author{Sin Kyu Kang}
\email{skkang@seoultech.ac.kr}
\affiliation{Institute of Convergence Fundamental Studies, Seoul National University
of Science and Technology, Seoul 01811, Korea} 
\affiliation{School of Natural Science, Seoul National University
of Science and Technology, Seoul 01811, Korea} 

\author{Anthony G. Williams}
\email{anthony.williams@adelaide.edu.au }
\affiliation{ARC Centre of Excellence for Dark Matter Particle Physics, Department of Physics, University
of Adelaide, South Australia 5005, Australia}

\date{\today}

\begin{abstract}
Primordial black holes (PBHs), hypothesized to form in the early universe from gravitational collapse of density fluctuations, represent a well-motivated dark matter (DM) candidate. Their potential detection through gamma-ray signatures arising from Hawking radiation would provide definitive evidence for their existence and constrain their contribution to the DM abundance. Unlike conventional DM candidates, PBHs emit a unique, thermal-like spectrum of particles as they evaporate, including photons, neutrinos, and possible beyond-the-Standard Model particles. Future high-sensitivity gamma-ray observatories, such as e-ASTROGAM and other next-generation telescopes, will play a pivotal role in this search. With improved energy resolution and sensitivity, these missions can disentangle PBH-originating photons from astrophysical backgrounds, probe subtle spectral features such as multi-peak structures, and test exotic evaporation models. Such observations could either confirm PBHs as a viable DM component or place stringent limits on their abundance across critical mass windows. In this work, we explore the distinguishing features of a double-peaked gamma-ray spectrum produced by PBHs, focusing on the asteroid-mass window ($10^{15}$ g to $10^{17}$ g), where Hawking radiation peaks in the MeV to GeV range. Using a likelihood-based analysis, we demonstrate how future missions could discriminate between single- and double-peaked PBH scenarios, the latter arising in cosmological models predicting multi-modal PBH mass distributions. Our results highlight the diagnostic power of spectral shape analysis in identifying PBH populations and constrain the parameter space for which a double-peaked signal could be detectable above background.

\end{abstract}

\maketitle
\section{Introduction}\label{Sec.1}

Primordial black holes (PBHs) represent a compelling macroscopic candidate for dark matter (DM), potentially comprising either the entirety or a portion of the cold DM relic abundance~\cite{Hawking:1971ei,Chapline:1975ojl,Khlopov:2008qy,Carr:2016drx,Carr:2020gox,Carr:2020xqk,Green:2020jor}. Their formation in the early universe can be attributed to several mechanisms: 
(i) the gravitational collapse of overdense regions seeded by primordial fluctuations generated during inflation~\cite{Carr:1974nx,Sasaki:2018dmp,Cheung:2023ihl}, 
(ii) energy concentration within the Schwarzschild radius resulting from bubble wall collisions during a first-order phase transition~\cite{Hawking:1982ga,Kodama:1982sf,Moss:1994iq,Konoplich:1999qq}, and 
(iii) the involvement of dark sector particles, where PBHs emerge from the collapse of large-scale intermediate structures known as fermi-balls~\cite{Baker:2021nyl,Gross:2021qgx,Kawana:2021tde,Marfatia:2021hcp}. 
For PBHs with masses below $10^{-15}$ solar masses, Hawking radiation becomes appreciable~\cite{Hawking:1975vcx,Gibbons:1977mu}, leading to the emission of both Standard Model (SM) and beyond the Standard Model (BSM) particles~\cite{Bell:1998jk,Mazde:2022sdx,Arbey:2021ysg,Masina:2021zpu,Schiavone:2021imu,Bernal:2021yyb,Mazde:2022sdx,Carr:1976zz,Toussaint:1978br,Turner:1979bt,Baumann:2007yr,Fujita:2014hha,Hook:2014mla,Hooper:2020otu,Hamada:2016jnq,Cheung:2025gdn}.

Studying the formation and evolution of PBHs provides a powerful tool for probing the early universe. The concept of PBHs was first proposed by Zel’dovich and Novikov~\cite{Zeldovich:1967lct}, and later developed by Hawking and Carr~\cite{Hawking:1971ei,Carr:1974nx,Carr:1975qj}, who suggested that such Black holes (BH) could have formed shortly after the Big Bang. According to theoretical predictions, PBHs can form in regions of the early universe where density perturbations exceed a critical threshold. A key motivation for investigating PBHs is their potential role as a natural DM candidate. Although recent observational constraints have placed stringent limits on their abundance, a viable mass window remains—specifically between $10^{16}~\mathrm{g}$ and $10^{20}~\mathrm{g}$—within which PBHs could still constitute a significant fraction of DM.

According to the conventional framework of PBHs evaporation, any PBHs with masses below $10^{15}$~g should have completely evaporated by now. Since the Hawking temperature increases as the BH mass decreases, the resulting photon spectrum typically peaks around 100~MeV for such evaporating PBHs.

However, recent theoretical developments~\cite{Dvali:2018xpy,Dvali:2020wft,Dvali:2024hsb} propose that the evaporation process may be significantly altered when one accounts for the back-reaction of the emitted radiation on the BH’s internal quantum state. This phenomenon, termed the \textit{memory burden} effect, arises from the quantum information retained by the BH, which tends to resist the loss of mass through radiation. As a BH sheds a substantial portion of its original mass, this effect can become strong enough to decelerate or even inhibit further evaporation, effectively extending the BH's lifespan. Under this scenario, PBHs with masses as low as $10^9$~g could potentially still exist today, and their associated photon emission would be shifted toward higher energies due to their smaller mass and consequently higher Hawking temperature.

Certain inflationary models predict the generation of double-peaked or multi-peaked power spectra by incorporating multiple inflection points or small features—such as bumps or dips—in the inflaton potential~\cite{ZhengRuiFeng:2021zoz,Zhang:2021vak}. Such models naturally give rise to a spectrum of PBHs with multiple mass peaks, as well as a gravitational wave (GW) spectrum featuring double-peaked structures. To explore the phenomenological implications, one can consider a toy model for the curvature power spectrum motivated by these scenarios: a lognormal double-peaked power spectrum given by
\begin{equation}
\label{Eq:1}
 \mathcal{P}_R = P_0 \ln\left(\frac{k}{k_f}\right)^2
 +A_1 \frac{\exp\left[-\ln\left(k/k_{p1}\right)^2/2 \sigma_0^2\right]}{\sqrt{2 \pi} \sigma_0}
 +A_2 \frac{\exp\left[-\ln\left(k/k_{p2}\right)^2/2 \sigma_0^2\right]}{\sqrt{2 \pi} \sigma_0} .
\end{equation}
This model consists of three main components. The first term, proportional to $P_0 \ln^2(k/k_f)$, represents a smooth logarithmic rise in the power spectrum, which can be interpreted as a background contribution. The second and third terms correspond to two lognormal peaks centered at comoving scales $k_{p1}$ and $k_{p2}$, respectively, each characterized by amplitude coefficients $A_1$ and $A_2$, and a common width parameter $\sigma_0$. These peaks model localized enhancements in the power spectrum due to specific features—such as bumps or inflection points—in the inflaton potential. Such enhancements are responsible for the formation of PBHs at different mass scales and the associated generation of GW.

% \begin{figure}
%     \begin{center}
%     \includegraphics[height=0.75\textwidth, width=\textwidth]{Figures/Toy_Model/Toy1.pdf}
%     \end{center}
%     \caption{
% Illustration of the toy model. 
% \textit{Top left:} The blue line represents the double-peaked power spectrum from \Eq{Eq:1}, calculated using the parameters in \Eq{Eq:2}. The CMB scale is indicated by a blue star, and the dashed red line marks the CMB normalization value of the power spectrum, $\mathcal{P}_R = 2.1 \times 10^{-9}$. 
% \textit{Top right:} The PBH DM fraction corresponding to the power spectrum, shown together with current observational constraints. 
% \textit{Bottom center:} The induced GW spectrum, $\Omega_{\rm GW} h^2(f)$, is compared with present (shaded regions) and future (red solid lines) experimental sensitivities.
% }
% 
%     \label{fig:Motivation}
% \end{figure}

\begin{figure}
  \centering
  \begin{subfigure}{0.49\textwidth}
    \centering
    \includegraphics[width=\linewidth]{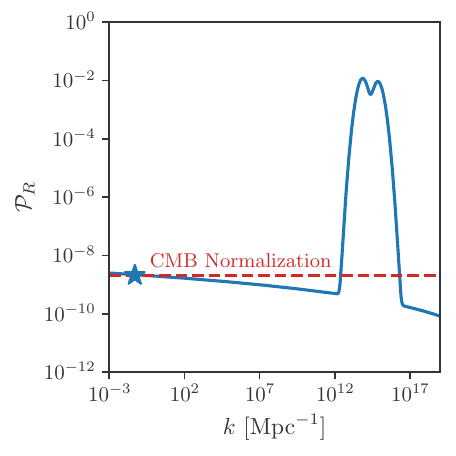}
  \end{subfigure}
  \begin{subfigure}{0.49\textwidth}
    \centering
    \includegraphics[width=\linewidth]{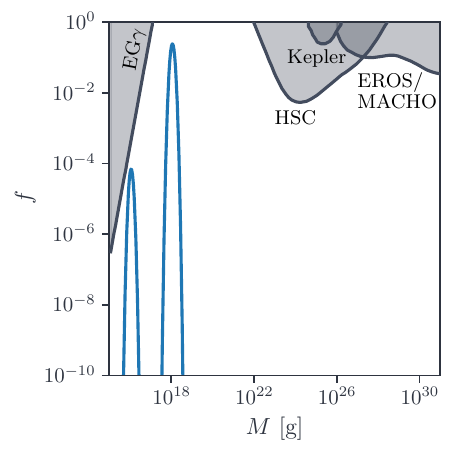}
  \end{subfigure}
  \begin{subfigure}{0.49\textwidth}
    \centering
    \includegraphics[width=\linewidth]{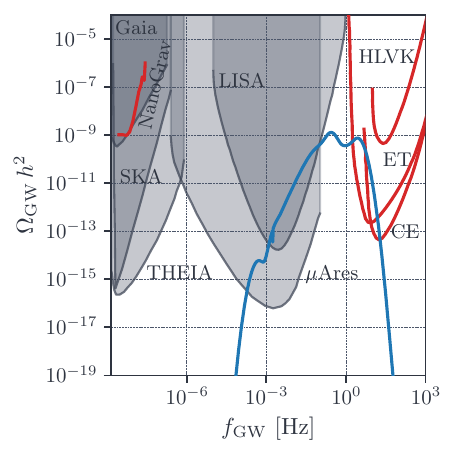}
  \end{subfigure}
  \caption{Illustration of the toy model. 
\textit{Top left:} The blue curve represents the double-peaked power spectrum from \Eq{Eq:1}, calculated using the parameters in \Eq{Eq:2}. The CMB scale is indicated by a blue star, and the dashed red line marks the CMB normalization value of the power spectrum, $\mathcal{P}_R = 2.1 \times 10^{-9}$. 
\textit{Top right:} The PBH DM fraction corresponding to the power spectrum, shown together with current observational constraints. 
\textit{Bottom center:} The induced GW spectrum, $\Omega_{\rm GW} h^2( f_{\rm GW})$, is compared with present (shaded regions) and future (red solid lines) experimental sensitivities.}
  \label{fig:Motivation}
\end{figure}

Using the double-peaked curvature power spectrum introduced above, one can quantitatively estimate both the PBHs abundance and the stochastic GW background generated at second order in perturbation theory. Consider the illustrative parameter choice,
\begin{equation}
\label{Eq:2}
\begin{aligned}
P_0 &= 6.57 \times 10^{-13}, \quad k_f = 7.7 \times 10^{23}~\mathrm{Mpc}^{-1}, \quad k_{p1} = 7.15 \times 10^{14}~\mathrm{Mpc}^{-1},\\ \quad k_{p2} &= 7.15 \times 10^{13}~\mathrm{Mpc}^{-1},
A_1 = 0.014, \quad A_2 = 0.018, \quad \sigma_0 = 0.6.
\end{aligned}
\end{equation}
This double lognormal spectrum exhibits two prominent peaks, leading to the production of PBHs at two distinct mass scales. The chosen parameters yield a power spectrum consistent with the observed value of approximately $2.1 \times 10^{-9}$ at the CMB scale ($k = 0.05\, \mathrm{Mpc}^{-1}$) while also producing peaks at the desired $k$ values.

The PBH abundance is determined by the mass fraction $\beta(M)$ at the time of formation, which is given by \cite{Niemeyer:1999ak,Musco:2004ak,Harada:2013epa,Press:1973iz,
Garcia-Bellido:2017mdw,Garcia-Bellido:1996mdl}
\begin{equation}
\label{Eq:3}
\beta(M) = 2 \int_{\delta_c=0.414}^{\infty} d\delta~P(\delta), \quad P(\delta) = \frac{1}{\sqrt{2\pi}\sigma(M)}\exp\left[-\left(\frac{\delta^2}{\sigma^2(M)}\right)\right],
\end{equation}
where the variance $\sigma^2(M)$ of the density contrast smoothed on the comoving scale $R$ is calculated as
\begin{equation}
\label{Eq:4}
\sigma^2(M) = \frac{4(1+w)^2}{(5+3w)^2} \int \frac{dk}{k} (kR)^4 W^2(k, R) \mathcal{P}_R(k), \quad W(k, R) = \exp\left[-\frac{k^2 R^2}{2}\right].
\end{equation}
The present-day PBH fraction relative to the DM density, $f_{\text{PBH}}$, is related to $\beta(M_{\text{PBH}})$ through the following expression~\cite{Niemeyer:1999ak,Musco:2004ak,Harada:2013epa,Press:1973iz,
Garcia-Bellido:2017mdw,Garcia-Bellido:1996mdl}:
\begin{equation}
\label{Eq:5}
\beta(M_{\text{PBH}}) = 3.7 \times 10^{-9} \left( \frac{\gamma}{0.2} \right)^{-1/2} \left( \frac{g_{* \text{form}}}{10.75} \right)^{1/4} \left( \frac{M_{\text{PBH}}}{M_\odot} \right)^{1/2} f_{\text{PBH}},
\end{equation}
where $\gamma$ is the efficiency factor for collapse and $g_{*\text{form}}$ is the number of relativistic degrees of freedom at PBH formation.

Moreover, the second-order induced GW resulting from enhanced curvature perturbations can be computed via the following integral~\cite{Baumann:2007zm,Di:2017ndc,Espinosa:2018eve,Kohri:2018awv,Inomata:2019yww,Pi:2024jwt,Halkoaho:2022hkp}:
\begin{align}
\label{Eq:6}
\Omega_\text{GW,eq}(k)
&= 3 \int_0^\infty \mathrm{d}v \int_{|1-v|}^{1+v} \mathrm{d}u \, \frac{1}{4u^2 v^2}
\left[ \frac{4v^2 - (1 + v^2 - u^2)^2}{4uv} \right]^2 \left( \frac{u^2 + v^2 - 3}{2uv} \right)^4 \nonumber \\
&\quad \cdot \left[ \left( \ln \left| \frac{3 - (u + v)^2}{3 - (u - v)^2} \right| - \frac{4uv}{u^2 + v^2 - 3} \right)^2 + \pi^2 \Theta \left( u + v - \sqrt{3} \right) \right] \nonumber \\
&\quad \cdot \left( \mathcal{P}_{\mathcal{R}}(uk) \right) \cdot \left( \mathcal{P}_{\mathcal{R}}(vk) \right).
\end{align}
where $\mathcal{P}_{\mathcal{R}}(k)$ denotes the gauge-invariant curvature perturbation power spectrum. The double-peaked structure in $\mathcal{P}_{\mathcal{R}}(k)$ leads to corresponding features in the GW spectrum, potentially observable by future experiments. The resulting curvature power spectrum, PBH abundance, and induced GW spectrum corresponding to the parameter choices encoded in \Eq{Eq:2} are illustrated in Fig.~\ref{fig:Motivation}. The \textit{top left} panel shows the double-peaked power spectrum as defined in \Eq{Eq:1}, evaluated using the parameters from \Eq{Eq:2}, with the CMB scale marked by a blue star. The \textit{top right} panel displays the resulting fraction of PBHs contributing to the DM abundance, along with current observational constraints shown as solid curves with shaded regions. These include extragalactic gamma-ray (EG$\gamma$) observations\cite{Carr:2009jm}, Subaru HSC microlensing results~\cite{Niikura:2017zjd}, Kepler milli/microlensing data~\cite{Griest:2013esa}, and EROS/MACHO microlensing bounds~\cite{EROS-2:2006ryy}. The \textit{bottom center} panel shows the GW energy density $\Omega_{\rm GW} h^2$ as a function of frequency $f$, derived from the curvature power spectrum. This is compared with the NANOGrav 15-year data~\cite{NANOGrav:2023ctt} and projected sensitivities of future experiments. These include pulsar timing arrays such as SKA and THEIA~\cite{Janssen:2014dka,Theia:2017xtk}, which are sensitive to stochastic backgrounds at $\mathcal{O}({\rm nHz})$, as well as space-based interferometers like LISA and $\mu$Ares~\cite{Caprini:2015zlo,Auclair:2019wcv,Sesana:2019vho} targeting the $\mu$Hz to Hz range. It is clearly evident from Fig.~\ref{fig:Motivation} that a double-peaked feature in the power spectrum can lead to the formation of PBHs at distinct mass scales, accompanied by a stochastic GW background with a characteristic double-peak structure.

\par In this study, we focus on the gamma-ray spectrum produced by a population of PBHs characterized by a double mass peak. Our aim is to explore the parameter space in which a double-peaked gamma-ray spectrum can be distinguished from both single-peaked spectra and astrophysical backgrounds. Such a distinctive feature can originate from the Hawking evaporation of PBHs formed at multiple mass scales. These PBHs are seeded by a double-peaked curvature power spectrum, which may, in turn, arise from inflationary scenarios featuring multiple inflection points or localized features—such as bumps or dips—in the inflaton potential~\cite{ZhengRuiFeng:2021zoz,Zhang:2021vak}. 

\par We analyze the gamma-ray spectrum with a double-peaked structure originating from PBHs within a $|R| < 5^{\circ}$ region around the Galactic Center, considering two distinct scenarios. These are: (i) the {\it SM-Case} (Single Peak), where gamma rays are emitted from a single PBH population, resulting in a spectrum with a single peak; and (ii) the {\it BSM-Case} (Double Peak), where gamma rays are produced from two distinct PBH populations with different masses, leading to a gamma-ray spectrum featuring two separate peaks.

In the SM-Case, the photon spectrum consists of contributions from various channels such as primary photons, neutral pion decays, final-state radiation (FSR) from electrons, and decay plus FSR from muons, among others~\cite{Arbey:2021mbl}.

Conversely, the BSM-Case introduces two distinct peaks in the gamma-ray spectrum, arising from PBHs of two different mass scales, denoted by $\mathrm{M_1}$ and $\mathrm{M_2}$. For this analysis, we consider PBH masses within the range $[10^{15}\,{\rm g},10^{17}\,{\rm g}]$ and employ the projected sensitivity of the e-ASTROGAM experiment~\cite{e-ASTROGAM:2016bph,Agashe:2022jgk} to conduct a likelihood-based statistical analysis. This allows us to delineate the regions of parameter space where the BSM Case can be statistically distinguished from the SM Case.

This work is structured as follows. In Section~\ref{Sec.2}, we review the gamma-ray spectrum arising from Hawking radiation in the double-peaked case and derive the integrated photon flux from the Galactic Center. Section~\ref{Sec.3} applies the future sensitivity of e-ASTROGAM to perform a statistical analysis aimed at identifying and assessing the discovery prospects for both the SM and BSM scenarios. Finally, we present our conclusions in Section~\ref{Sec.4}.

\section{Particle Emission via Hawking Radiation from Primordial Black Holes}\label{Sec.2}

\begin{figure}[t]
    \centering
    \includegraphics[width=0.98\textwidth]{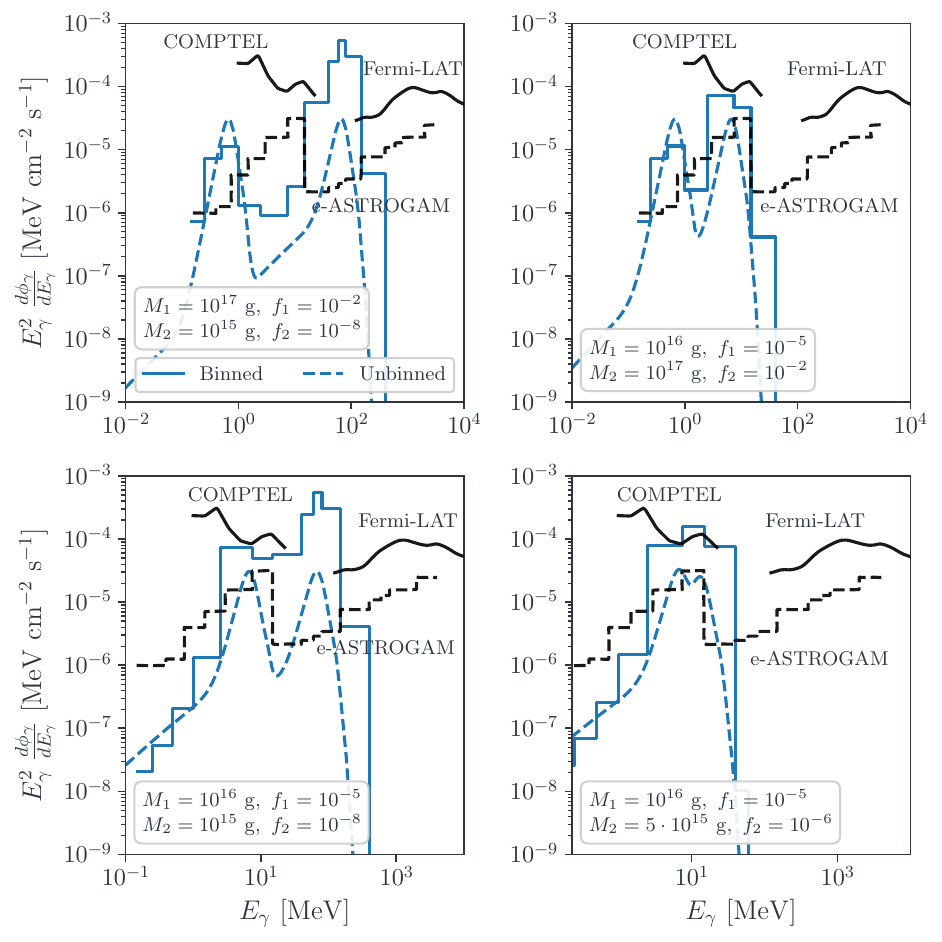}
    \caption{
        Double-peaked gamma-ray spectra from Hawking radiation of PBHs for different mass and abundance combinations (labeled in each panel). The spectra are shown as both binned (solid blue) and unbinned (dashed blue) versions of the same underlying model. %The region of interest is the Galactic center with \(|R| \leq 5^\circ\). 
        Current constraints and future sensitivities from COMPTEL (solid), Fermi-LAT~\cite{Fermi-LAT:2018pfs} (solid), and e-ASTROGAM~\cite{e-ASTROGAM:2016bph,Agashe:2022jgk} (dashed) are also shown.
    }
    \label{fig:spectrum}
\end{figure}

BHs are expected to emit particles continuously from their event horizon due to quantum effects, a process known as Hawking radiation~\cite{Hawking:1974rv}. In this section, we examine the spectrum of particles emitted through this mechanism~\cite{Coogan:2020tuf}. The particles released directly from the BH are referred to as \textit{primary particles}, while those arising from subsequent interactions and decays of the primaries are classified as \textit{secondary particles}. The differential emission rate for a primary particle species $i$ from a black hole of mass $M$, per unit time and energy, is given by~\cite{Hawking:1974rv, Page:1976df, MacGibbon:1990zk}:
\begin{equation}\label{eq:7}
    \frac{\partial N_{i,\textrm{primary}}}{\partial E_i \partial t} = \frac{g_i}{2 \pi} \frac{\Gamma_i(E_i,M,m_i)}{e^{E_i/T_H} \pm 1},
\end{equation}
where $m_i$ and $g_i$ denote the particle's mass and internal degrees of freedom, respectively. The quantity $\Gamma_i$ is the graybody factor, and $T_H = 1/(8 \pi G M)$ represents the Hawking temperature. The sign convention distinguishes fermions ($+$) from bosons ($-$).

The graybody factors describe the probability that a particle which generated at the horizon of a BH, propagate to spatial infinity. They are obtained by solving the particle's equation of motion in curved spacetime with specific boundary conditions. At high energies, the graybody factor approaches the geometrical optics limit: $\Gamma_i \rightarrow 27 G^2 M^2 E_i^2$. In our analysis, we employ the \texttt{BlackHawk v2.0} software package~\cite{Arbey:2019mbc, Arbey:2021mbl} to compute $\Gamma_i$ for non-rotating BHs. 
\texttt{BlackHawk v2.0} incorporates the effects of particle masses and introduces a natural cutoff in the emission spectrum for $E_i < m_i$, suppressing the production of particles with rest mass exceeding the BH's temperature.

Our primary focus is on the photon spectrum resulting from Hawking radiation, which includes both direct emissions (primary photons) and secondary photons produced through decays and final-state radiation (FSR) of heavier particles. The total photon emission spectrum is given by:
\begin{eqnarray}\label{eq:8}
\frac{\partial N_{\gamma,\textrm{tot}}}{\partial E_\gamma \partial t} &=& \frac{\partial N_{\gamma,\textrm{primary}}}{\partial E_\gamma \partial t} 
+ \int d E_{\pi^{0}}~2\frac{\partial N_{\pi^{0},\textrm{primary}}}{\partial E_{\pi^{0}} \partial t} \frac{d N_{\pi^{0},\textrm{decay}}}{dE_\gamma} \nonumber \\
&& + \sum_{i=e^\pm,\mu^\pm,\pi^\pm} \int d E_i \frac{\partial N_{i,\textrm{primary}}}{\partial E_i \partial t} \frac{d N_{i,\textrm{FSR}}}{dE_\gamma}.
\end{eqnarray}
The relevant decay and radiation distributions are defined as follows:
\begin{eqnarray}\label{eq:9}
    \frac{d N_{\pi^0,\textrm{decay}}}{dE_\gamma} &=& \frac{\Theta(E_\gamma - E_{\pi^0}^-) \Theta(E_{\pi^0}^+ - E_\gamma)}{E_{\pi^0}^+ - E_{\pi^0}^-},\\\label{eq:10}
    E_{\pi^0}^\pm &=& \frac{1}{2} \left( E_{\pi^0} \pm \sqrt{E_{\pi^0}^2 - m_{\pi^0}^2} \right),
\end{eqnarray}
\begin{eqnarray}\label{eq:11}
    \frac{d N_{i,\textrm{FSR}}}{dE_\gamma} &=& \frac{\alpha}{\pi Q_i} P_{i\rightarrow i\gamma}(x) \left[ \log\left(\frac{1 - x}{\mu_i^2}\right) - 1 \right],\\\label{eq:12}
    P_{i\rightarrow i\gamma}(x) &=& 
    \begin{cases}
        \frac{2(1 - x)}{x}, & \text{for } i = \pi^\pm, \\
        \frac{1 + (1 - x)^2}{x}, & \text{for } i = \mu^\pm, e^\pm,
    \end{cases}
\end{eqnarray}
where $x = 2E_\gamma / Q_i$, $\mu_i = m_i / Q_i$, and $Q_i = 2E_i$. 

The differential photon flux observable near Earth from evaporating PBHs is given by\footnote{For simplicity, we consider a monochromatic mass distribution for PBHs. In the case of a double-peaked gamma-ray spectrum originating from two distinct PBH populations with different masses, the total photon flux is computed by evaluating the flux for each population separately under the monochromatic assumption and then summing the contributions to obtain the overall flux. 
}:
\begin{equation}\label{eq:13}
    \frac{d \Phi_\gamma}{d E_\gamma} = \bar{J}_D \frac{\Delta \Omega}{4 \pi} \int dM\, \frac{f(M)}{M} \frac{\partial N_{\gamma,\textrm{tot}}}{\partial E_\gamma \partial t},
\end{equation}
where $\bar{J}_D$ denotes the decay J-factor, defined as:
\begin{equation}\label{eq:14}
    \bar{J}_D = \frac{1}{\Delta \Omega} \int_{\Delta \Omega} d\Omega \int_{\textrm{LOS}} dl\, \rho_{\textrm{DM}}(r(l, \Omega)).
\end{equation}
Here, $\rho_{\textrm{DM}}(r)$ is the DM density profile, and the integration is performed along the line of sight (LOS) over a solid angle $\Delta \Omega$. For the Galactic DM distribution, we adopt the Navarro–Frenk–White (NFW) profile~\cite{Navarro:1996gj}:
\begin{equation}\label{eq:15}
\rho_{\rm DM}(r) = \frac{\rho_s}{(r/r_s) \left[1 + (r/r_s) \right]^2} \Theta(r_{200} - r),
\end{equation}
where the parameters are chosen as: scale radius $r_s = 11~\mathrm{kpc}$, characteristic density $\rho_s = 0.838~\mathrm{GeV}/\mathrm{cm}^3$, virial radius $r_{200} = 193~\mathrm{kpc}$, and the solar radius $r_\odot = 8.122~\mathrm{kpc}$~\cite{deSalas:2019pee}.

For a region of interest (ROI) centered on the Galactic Center within an angular radius of $|R| < 5^\circ$, the corresponding J-factor is $\bar{J}_D = 1.597 \times 10^{26}~\mathrm{MeV}\,\mathrm{cm}^{-2}\,\mathrm{sr}^{-1}$, with a solid angle $\Delta \Omega = 2.39 \times 10^{-2}\,\mathrm{sr}$.

\par A double-peaked gamma-ray spectrum, which can result from the evaporation of PBHs formed at two distinct mass scales, is illustrated in Fig.~\ref{fig:spectrum} for different combinations of \( \rm M_1 \) and \( \rm M_2 \). The energy scale corresponding to each peak is inversely proportional to the mass of the respective PBHs. To produce the binned spectra shown in the figure, we have adopted the same energy binning as reported in the experimental proposal~\cite{e-ASTROGAM:2016bph}. The values of \( f_{1}\) and \( f_{2} \) used in the plot are chosen from the allowed parameter space, as determined by the test statistic analysis performed to distinguish a double-peaked gamma-ray spectrum from a single-peaked one, as discussed in Sec.~\ref{Sec.4}.

\section{Results \& Discussion}\label{Sec.3}

\begin{figure}
    \begin{center}
    \includegraphics[width=0.85\textwidth]{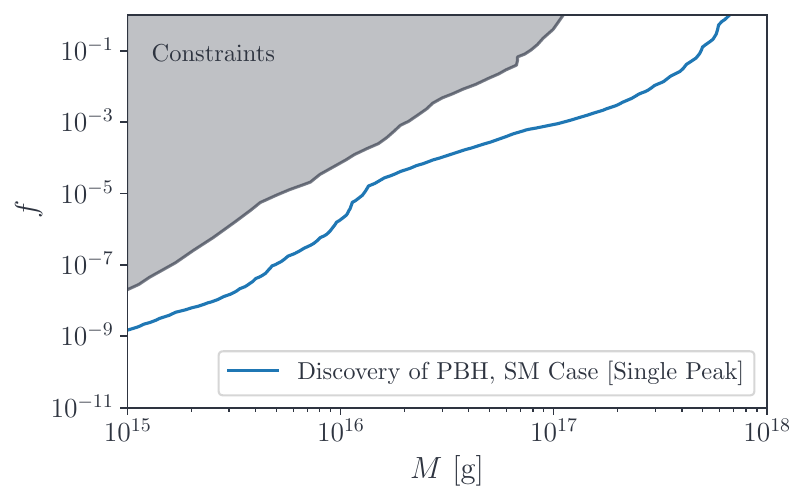}
    \end{center}
    \caption{Bounds on PBH detectability in the \( f \) \textit{vs.} \( \rm M \) plane. The blue curve (\( \mathrm{TS} = 9 \)) represents the lower bound on \( f \) required to distinguish a SM signal (a single peak) from astrophysical backgrounds. The gray shaded region indicates existing experimental constraints.
 }
    \label{fig:SM}
\end{figure}

%%%% Sen Fig1 %%%%
\begin{figure}
    \centering

    \begin{subfigure}[b]{0.49\textwidth}
        \includegraphics[width=\textwidth]{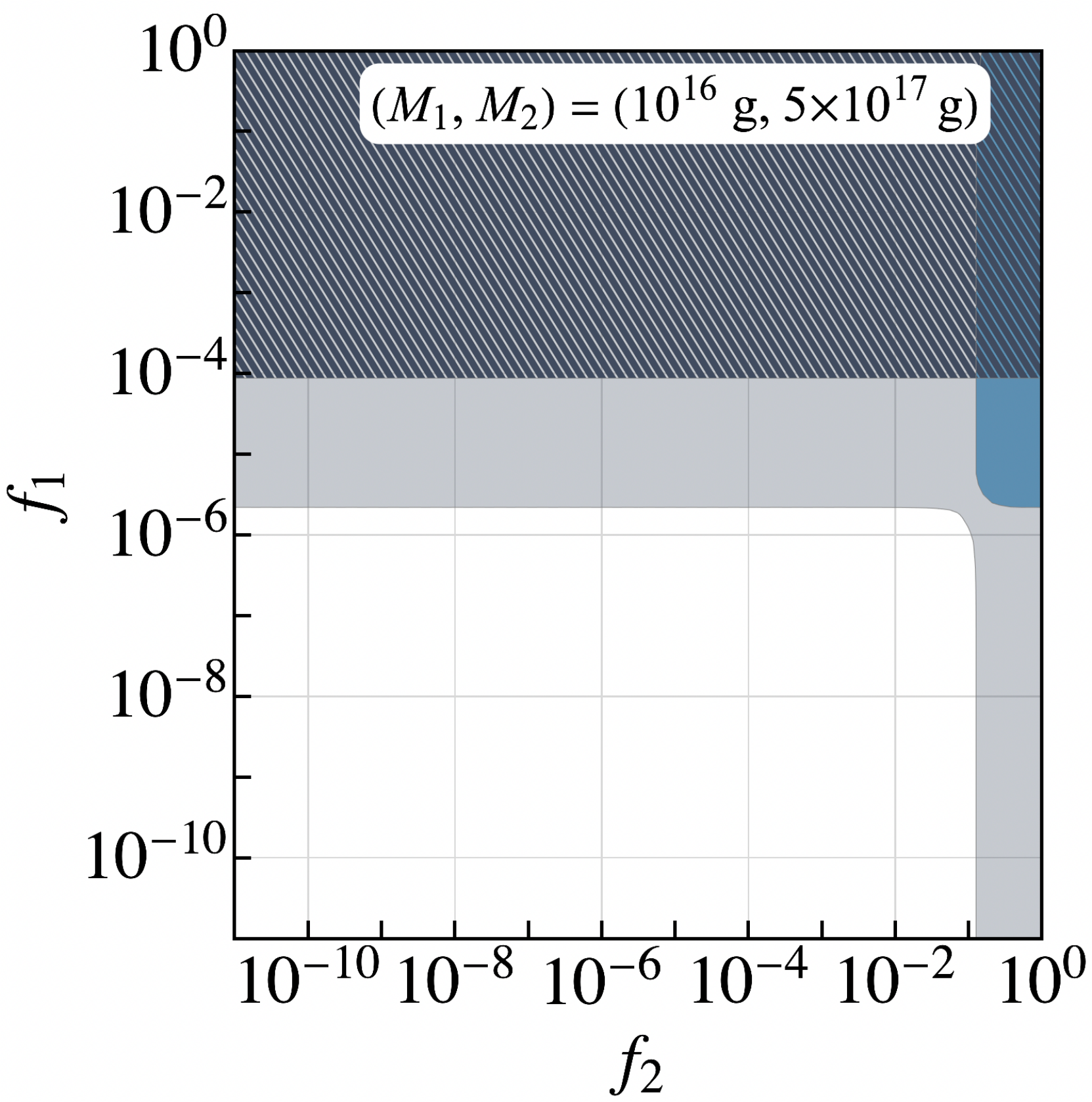}
    \end{subfigure}
    \hfill
    \begin{subfigure}[b]{0.49\textwidth}
        \includegraphics[width=\textwidth]{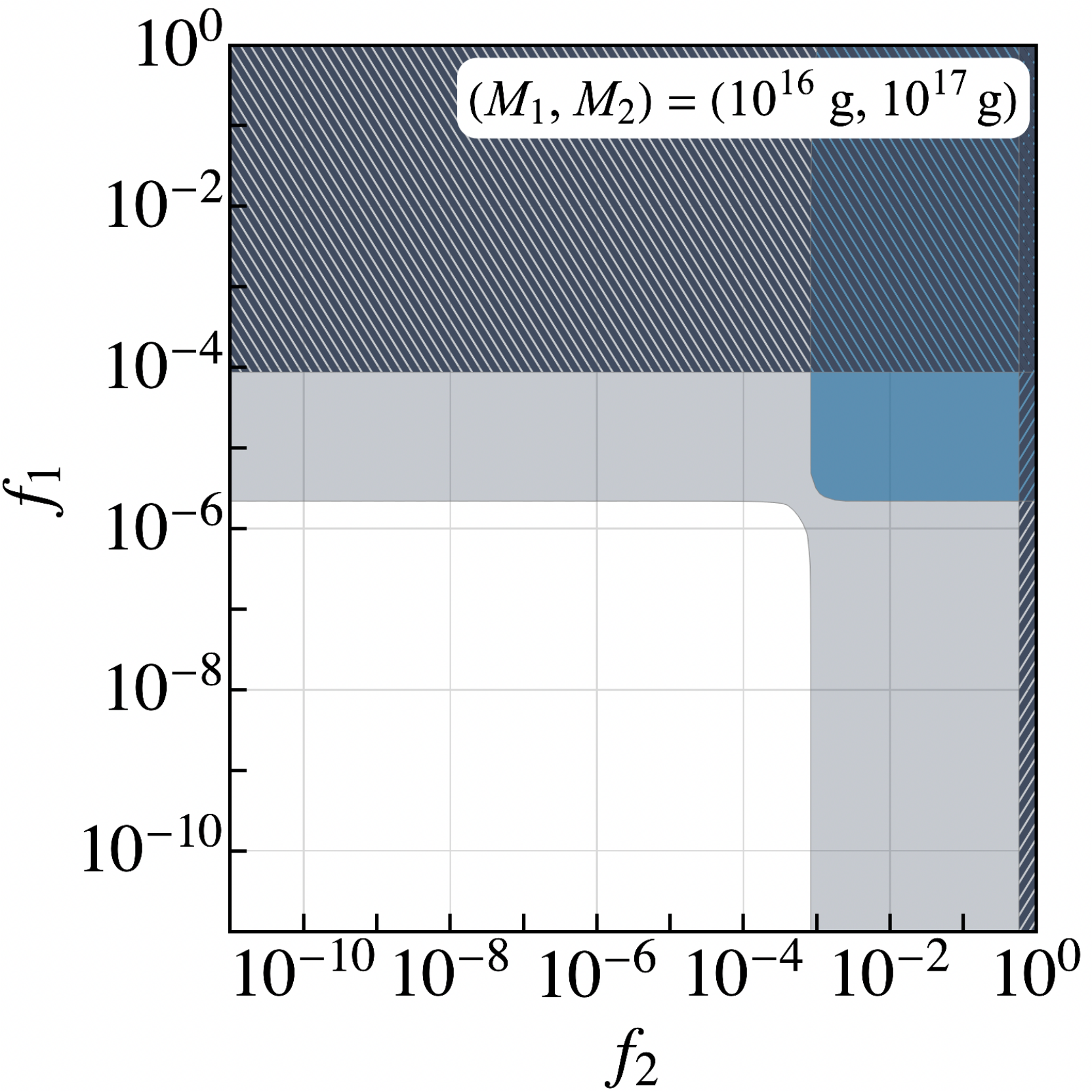}
    \end{subfigure}

\caption{
Constraints on PBH detectability in the \(f_2\) vs. \(f_1\) plane for fixed \(\rm M_1\) and \(\rm M_2\). The light gray and blue shaded regions represent the parameter space where a double-peaked gamma-ray spectrum can be distinguished from the astrophysical background. 
The blue region, in particular, denotes the subset where the double-peaked spectrum can also be distinguished from a single-peaked one. 
The white region corresponds to scenarios with no detectable PBH-induced gamma-ray signal. 
Gray hatched areas indicate existing observational constraints on the PBH mass.
}

\label{fig:sensitivity_1}
\end{figure}

%%%% Sen Fig2 %%%%
\begin{figure}
    \centering

    \begin{subfigure}[b]{0.49\textwidth}
        \includegraphics[width=\textwidth]{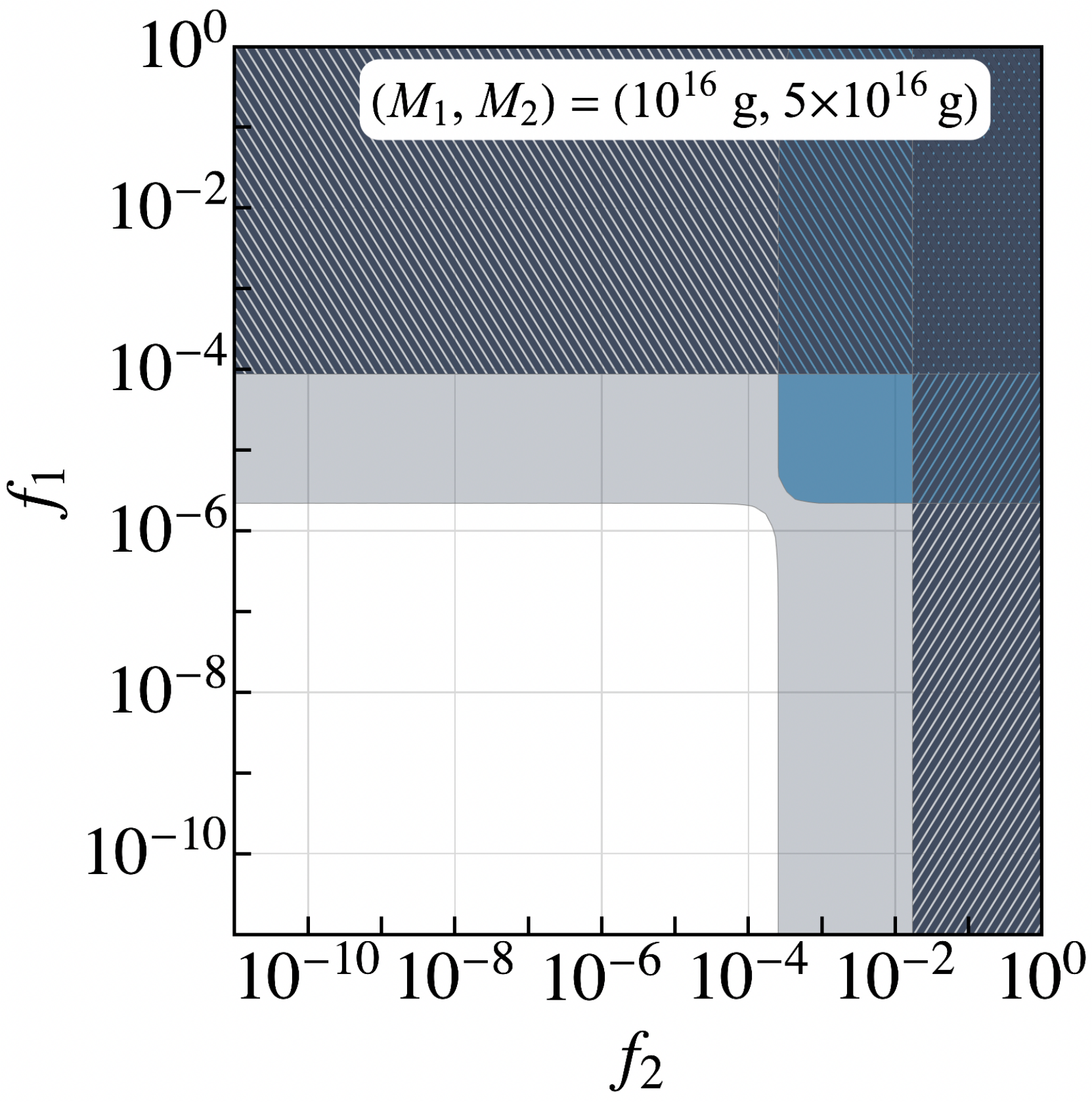}
    \end{subfigure}
    \hfill
    \begin{subfigure}[b]{0.49\textwidth}
        \includegraphics[width=\textwidth]{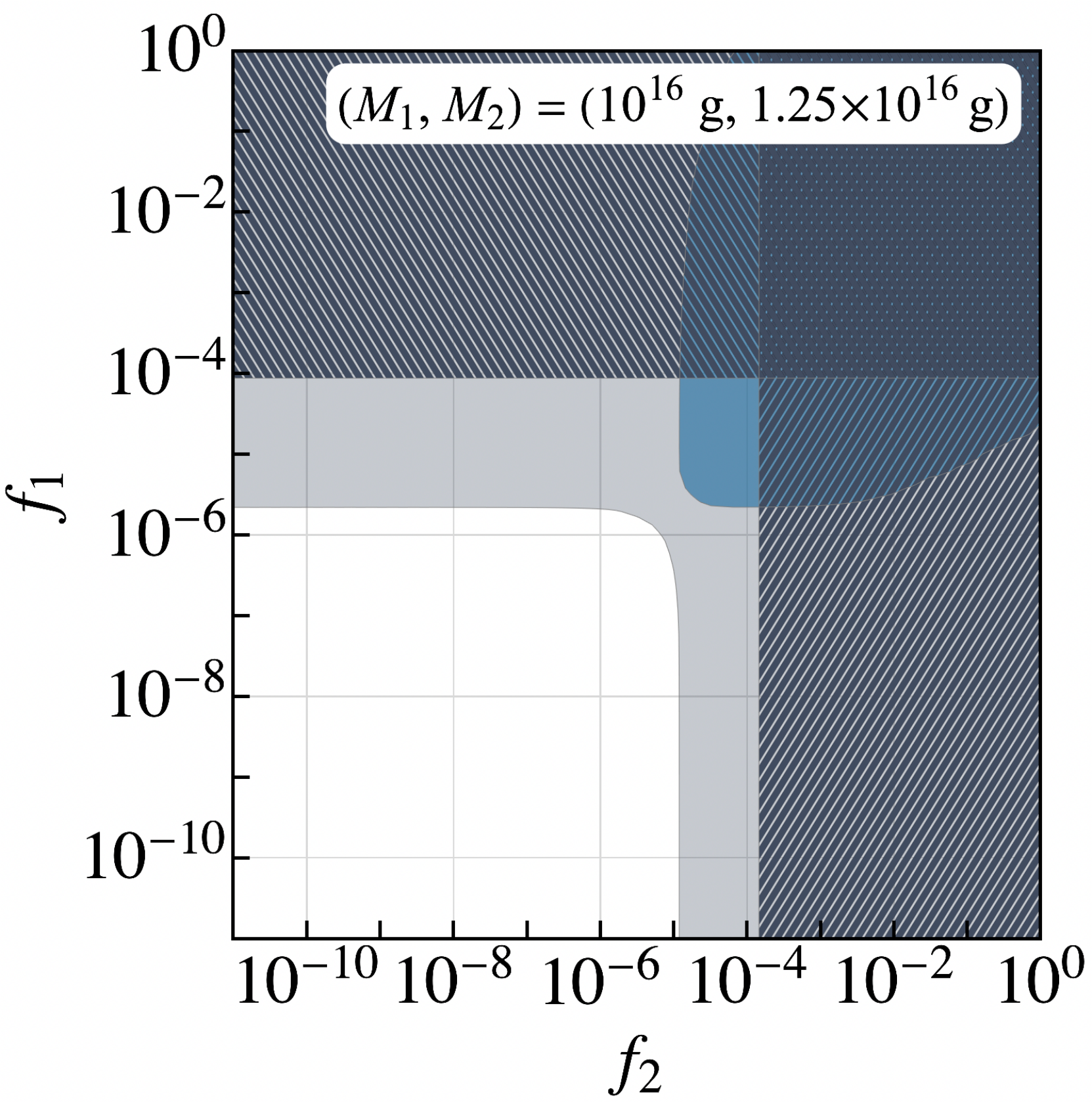}
    \end{subfigure}

    \caption{
Same as Fig.~\ref{fig:sensitivity_1}, but with different PBH mass combinations.}

    \label{fig:sensitivity_2}
\end{figure}

%%%% Sen Fig3 %%%%

\begin{figure}
    \centering

    \begin{subfigure}[b]{0.49\textwidth}
        \includegraphics[width=\textwidth]{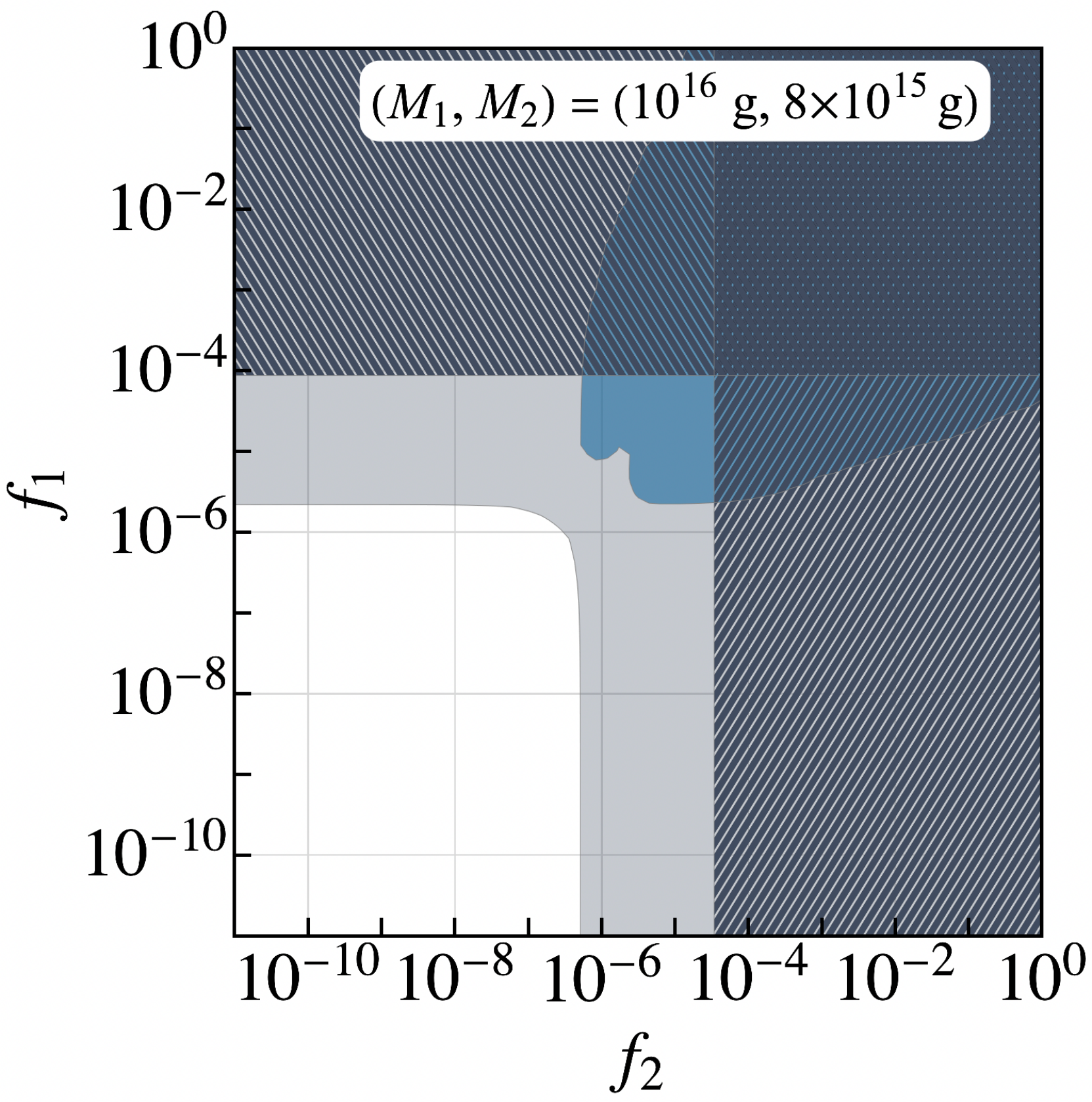}
    \end{subfigure}
    \hfill
    \begin{subfigure}[b]{0.49\textwidth}
        \includegraphics[width=\textwidth]{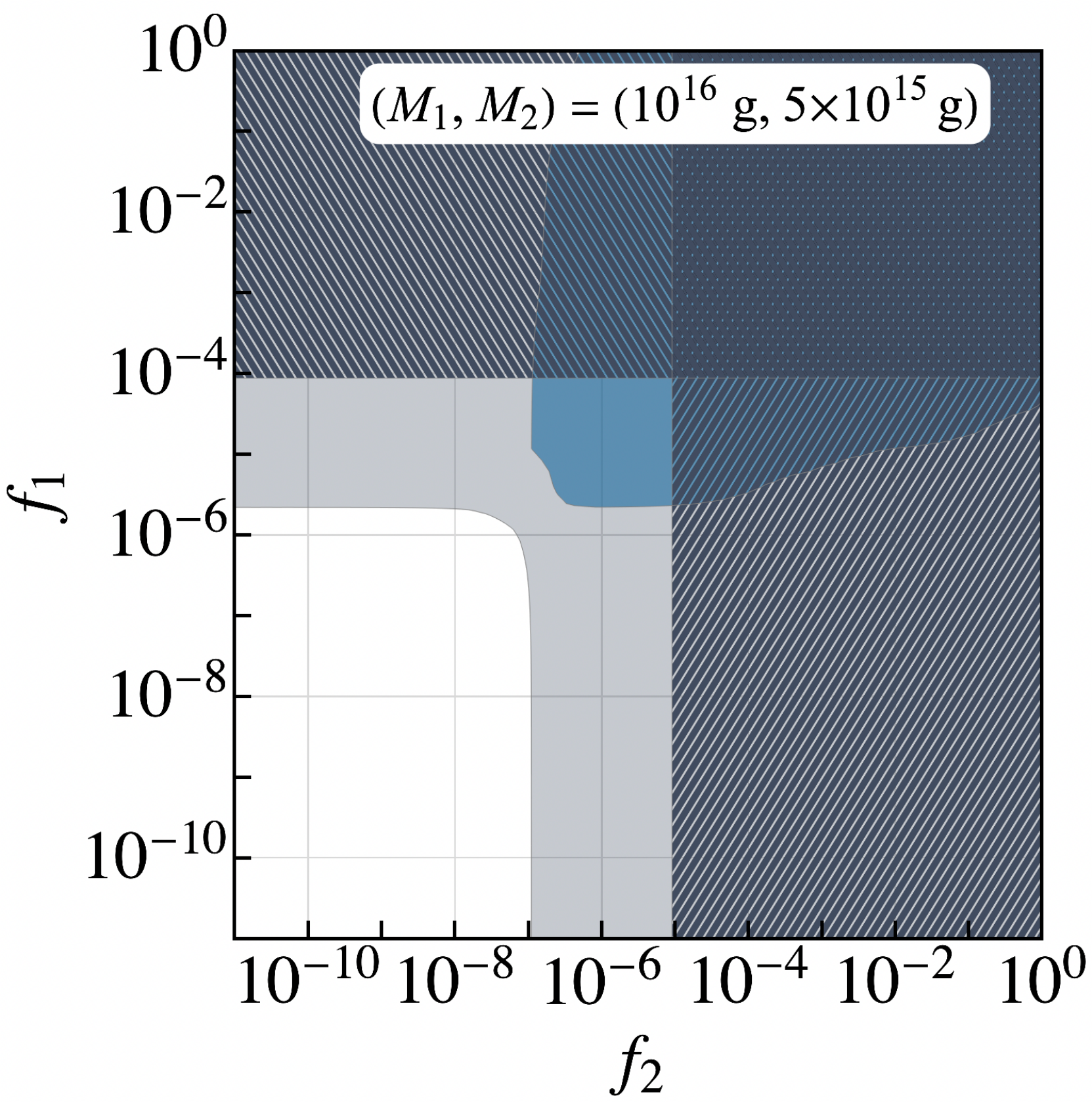}
    \end{subfigure}

    \caption{Same as Fig.~\ref{fig:sensitivity_1}, but with different PBH mass combinations.
    }
    \label{fig:sensitivity_3}
\end{figure}

%%%% Sen Fig4 %%%%

\begin{figure}
    \centering

    \begin{subfigure}[b]{0.49\textwidth}
        \includegraphics[width=\textwidth]{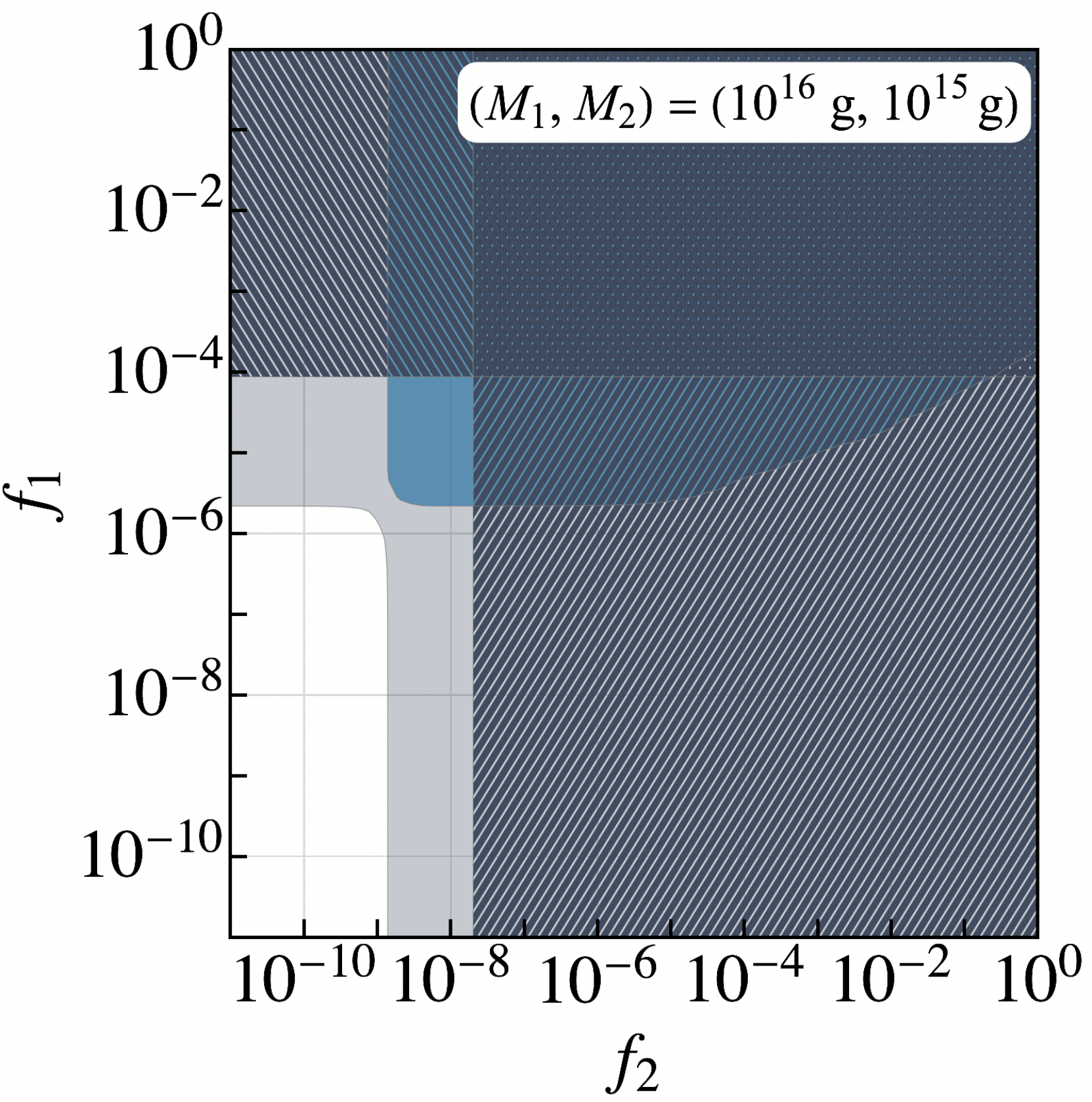}
    \end{subfigure}
    \hfill
    \begin{subfigure}[b]{0.49\textwidth}
        \includegraphics[width=\textwidth]{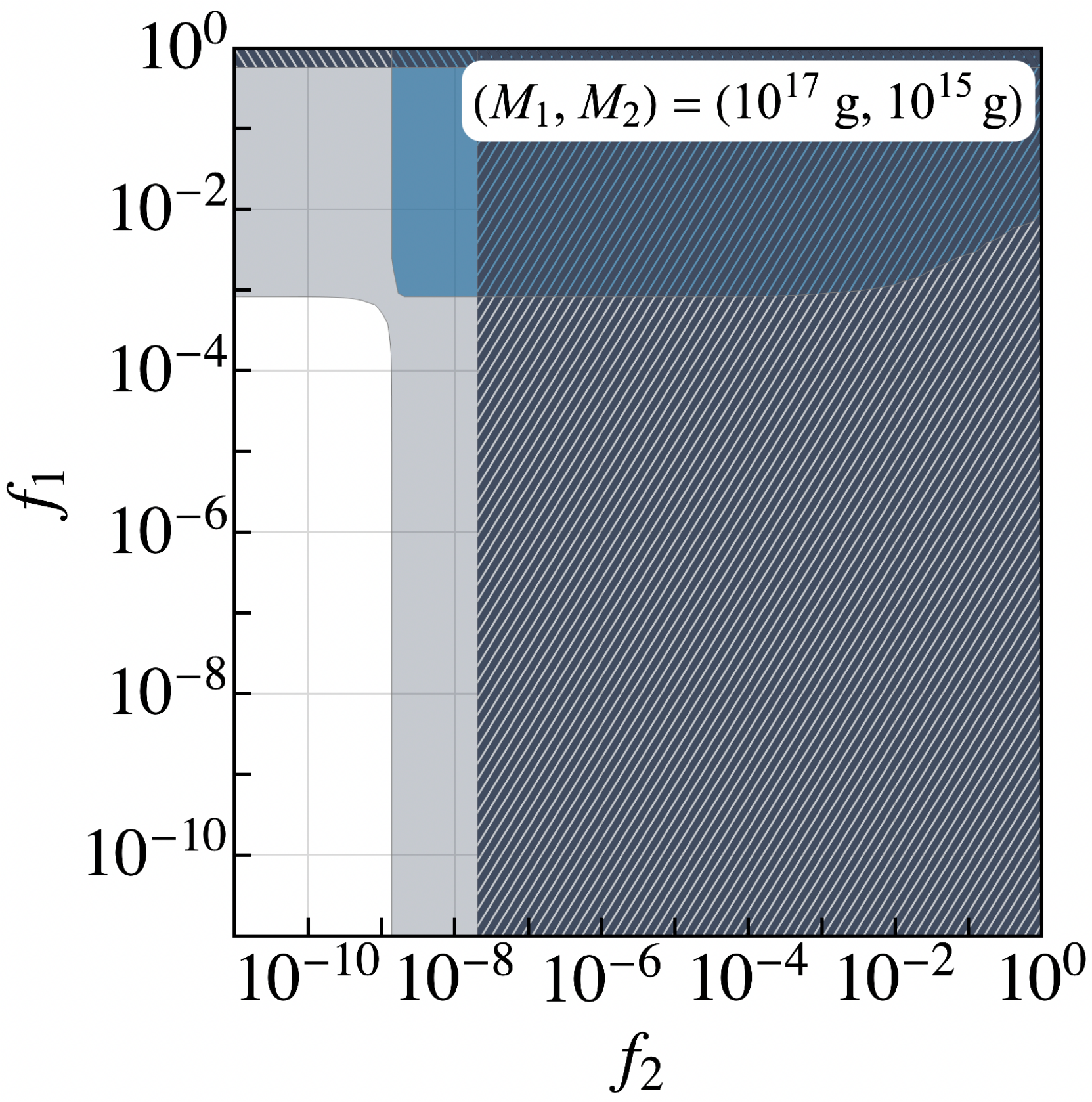}
    \end{subfigure}

    \caption{Same as Fig.~\ref{fig:sensitivity_1}, but with different PBH mass combinations. 
    }
    \label{fig:sensitivity_4}
\end{figure}

We employ a likelihood-based statistical analysis to constrain our model, 
using a True model that represents an observable gamma-ray signal 
as inferred from astrophysical backgrounds reported by experiments~\cite{Agashe:2022jgk,Agashe:2022phd}. 
A Test model incorporating new physics is then compared to the True model. 
In our study, we consider two Test models involving PBHs: (i) the SM scenario (Single-Peaked Case), 
and (ii) the BSM scenario (Double-Peaked Case).

The probability that a Test model reproduces the gamma-ray signal predicted by the True model 
is governed by Poisson statistics, and the likelihood is given by:
\begin{equation}
\mathcal{L}= \exp\left(\sum_i n_i \ln{\sigma_i} - \sigma_i - \ln{n_i!}\right),  
\label{eq.22}
\end{equation}  
where $n_i$ denotes the observed photon counts in the $i$-th energy bin, 
which includes the background contribution, and $\sigma_i$ is the expected photon count 
from the Test model (also including background) in the same bin\footnote{In this work, all results assume an
observation time of $10^8~\mathrm{s}$}.

To evaluate the relative compatibility of different Test models with the True model, 
we define the test statistic (TS) as:
\begin{equation}
    \mathrm{TS} = -2 \ln\left(\frac{\mathcal{L}}{\mathcal{L}_\mathrm{True}}\right) = \Sigma^2,  
    \label{eq:23}
\end{equation}
where $\Sigma$ denotes the significance level of the observation~\cite{Cowan:2010js, Rolke:2004mj, 
Bringmann:2012vr, Fermi-LAT:2015kyq}, and $\mathcal{L}_\mathrm{True}$ is the likelihood 
associated with the True model.

This analysis assumes that the cumulative statistical behavior across energy bins 
follows a $\chi^2$ distribution. Unless specified otherwise, we adopt a significance threshold 
of $\Sigma = 3$ for model exclusion or detection claims.

In this study, both the True and Test gamma-ray signals, along with the corresponding background estimates, are derived based on emission originating from the Galactic Center region. The DM distribution is modeled using the Navarro--Frenk--White (NFW) profile, assuming a $5^\circ$ observational cone around the Galactic Center. The analysis accounts for the projected sensitivity and detector response of upcoming gamma-ray instruments. Furthermore, the astrophysical background is incorporated to provide a realistic framework for evaluating the detectability of potential signals~\cite{e-ASTROGAM:2016bph, Agashe:2022jgk}.\footnote{Detailed descriptions of the instrumental sensitivity and background models can be found in Refs.~\cite{Agashe:2022jgk,Agashe:2022phd,e-ASTROGAM:2016bph}.}

\subsection{Discovery of a Single Peaked gamma ray spectrum~(SM Case)}

To investigate the identification of a single-peaked gamma-ray spectrum arising from the evaporation of a PBH with a single mass scale---a scenario extensively studied in the literature~\cite{Cheung:2025gdn,Agashe:2022jgk,Agashe:2022phd}---we consider the photon flux from a PBH of mass \( \rm M \) and abundance \( f \) as the Test model. The astrophysical background expected from the e-ASTROGAM detector is treated as the True model. We perform a likelihood-based statistical test, as described in the previous section, to determine the lower bounds on \( f \) as a function of PBH mass \( \rm M \). 

As previously discussed, we adopt a significance threshold of \( \mathrm{TS} = 9 \) to claim detection or exclusion in the SM scenario. The resulting constraints on PBH detectability in the \( f \) vs. \( \rm M \) plane are shown in Figure~\ref{fig:SM}. The blue curve corresponds to the \( \mathrm{TS} = 9 \) contour. Regions above this curve indicate detectable signals (\( \mathrm{TS} > 9 \)), while regions below it correspond to scenarios with no detectable PBH signal (\( \mathrm{TS} < 9 \)).

\subsection{Discovery and Identification of a Double-Peaked Gamma-Ray Spectrum (BSM Case)}

The double-peaked gamma-ray spectrum from PBHs arises from the superposition of emissions from two PBH populations with distinct masses, \(\mathrm{(M_1, M_2)}\). The positions of the spectral peaks depend on these masses, while the peak heights are governed by the PBH abundance fractions, \(f_1\) and \(f_2\). This subsection explores the \(f_1\) vs. \(f_2\) parameter space where a double-peaked spectrum can be distinguished from both the astrophysical background and a single-peaked PBH spectrum, using the future MeV gamma-ray experiment e-ASTROGAM.

To identify the parameter space where a double-peaked spectrum is distinguishable from the astrophysical background, we perform a TS analysis, as described in the previous section. Here, the Test model includes two PBH mass components, \(\mathrm{(M_1, M_2)}\), while the True model assumes only the background. Regions in the \(f_1\) vs. \(f_2\) parameter space satisfying \(\mathrm{TS} \geq 9\) are shown as light gray shaded areas in Figures~[\ref{fig:sensitivity_1}--\ref{fig:sensitivity_4}] for various \(\mathrm{(M_1, M_2)}\) combinations. These regions indicate where a double-peaked spectrum can be detected above the background.

For a given mass combination \(\mathrm{(M_1, M_2)}\), the Hawking radiation from the lighter PBH typically dominates the emission due to its faster evaporation rate. Since the total gamma-ray flux emitted by a PBH is proportional to its abundance \(f\) and inversely proportional to its mass, the relative contributions of each PBH population to the overall spectrum are highly sensitive to the chosen values of \( f_1 \) and \( f_2 \). Consequently, the height and intensity of each peak in the resulting double-peaked gamma-ray spectrum are directly governed by these abundance parameters. From an observational perspective, this implies that optimizing the values of \( f_1 \) and \( f_2 \) within the allowed parameter space is essential for maximizing the detectability of distinct spectral features, particularly in future MeV gamma-ray experiments such as e-ASTROGAM.

To distinguish a double-peaked spectrum from a single-peaked PBH spectrum, we compare a Test model (double-peaked with background) against a True model (single-peaked with background). This analysis yields a more constrained \(f_1\) vs. \(f_2\) parameter space, requiring larger abundance values for a given \(\mathrm{(M_1, M_2)}\) to achieve \(\mathrm{TS} \geq 9\). These identification regions are depicted as blue shaded areas in Figures~[\ref{fig:sensitivity_1}--\ref{fig:sensitivity_4}], with boundaries at \(\mathrm{TS} = 9\) marking the minimum \(f_1\) and \(f_2\) needed to differentiate the double-peaked signal from a single-peaked one. The gray hatched regions in Figures~[\ref{fig:sensitivity_1}--\ref{fig:sensitivity_4}] represent existing observational constraints on the corresponding PBH mass ranges.

When \(\mathrm{M_{1}} \approx \mathrm{M_{2}}\), the resulting gamma-ray peaks lie close in energy and eventually merge into a single peak as \(\mathrm{M_{1} = M_{2}}\). In such cases, the energy resolution of the detector becomes essential for distinguishing between a double- and single-peaked spectrum. For example, for \(\mathrm{(M_1, M_2)} = (10^{16}~\mathrm{g}, 5\times10^{15}~\mathrm{g})\), the peaks fall within the same energy bin, as shown in the bottom-right panel of Figure~\ref{fig:spectrum}. As the mass separation narrows, the spectrum increasingly resembles a single peak, making spectral resolution a critical factor in identifying distinct contributions. This effect is further illustrated in the right panel of Figure~\ref{fig:sensitivity_2}, corresponding to \(\mathrm{(M_1, M_2)} = (10^{16}~\mathrm{g}, 1.25\times10^{16}~\mathrm{g})\), where the distinguishable parameter space in the \(f_1\)\textit{–}\(f_2\) plane is tightly constrained. The boundary of the blue region corresponds to a TS value of 9, with higher TS values indicated by the shaded area. However, once observational constraints (gray hatched regions) are imposed, even these higher-TS regions become inaccessible for this mass configuration.
Regions with \(\mathrm{TS} < 9\), shown as white areas in Figures~[\ref{fig:sensitivity_1}--\ref{fig:sensitivity_4}], indicate no PBH detectability.

\section{Conclusion}\label{Sec.4}
PBHs remain a compelling DM candidate, offering a unique observational signature through their Hawking radiation, particularly in the form of gamma-ray emission. Focusing on the asteroid-mass range ($\rm 10^{15}~g-10^{17}~g$), we have demonstrated how a double-peaked gamma-ray spectrum could serve as a distinctive probe of multi-modal PBH mass distributions predicted in certain cosmological scenarios. Using a likelihood-based approach, we showed that future missions like e-ASTROGAM could effectively discriminate between single- and double-peaked spectra, providing a powerful test of PBH DM models.

The spectral shape of PBH evaporation not only offers a way to identify their presence but also opens a window into exotic physics, including potential BSM particle emissions. While future high-sensitivity gamma-ray telescopes will be crucial in resolving these signatures, challenges such as background modeling and peak separation must be addressed. Further research should explore extended mass distributions, additional particle emission channels, and multi-messenger correlations with GW data. This work establishes a foundation for leveraging next-generation gamma-ray observations to constrain PBH DM and uncover new insights into the early universe.

%\appendix
%\input{subtex/appendix1}

\section*{Acknowledgment}
C.J.O. extends sincere thanks to the ARC Centre of Excellence for Dark Matter Particle Physics and the Department of Physics, University of Adelaide, South Australia 5005, Australia, for their generous hospitality during the academic visit and the research conducted therein. C. J. O. and S. K. K are supported by the National Research Foundation of Korea under grant NRF-2023R1A2C100609111. GB, JG and AGW are funded by the ARC Centre of Excellence for Dark Matter Particle Physics CE200100008 and are further supported by the Centre for the Subatomic Structure of Matter (CSSM).

\bibliography{paper}

\end{document}